\PassOptionsToPackage{dvipsnames}{xcolor}
\documentclass[sigplan,nonacm]{acmart}
\setkeys{acmart.cls}{balance=false}
\setcopyright{none}
\hypersetup{pdftitle={Characterizing Performance Tradeoffs in Heterogeneous GPU--Wafer-Scale Accelerator Systems},pdfauthor={Ziyu Hu, Zhijing Ye, Jiamin Wang, Zhiqing Zhong, Zhendong Mi, Shaoyi Huang, Weijian Zheng, Rajkumar Kettimuthu, Xiaodong Yu}}

\usepackage[normalem]{ulem}
\usepackage{graphicx}
\usepackage{textcomp}
\usepackage{xcolor}
\usepackage{booktabs}
\usepackage{array}
\usepackage{tabularx}
\usepackage{longtable}
\usepackage{multirow}
\usepackage{url}
\usepackage{enumitem}
\usepackage{tcolorbox}
\usepackage{needspace}

\usepackage{libertine}

\begin{document}
\raggedbottom

\title[GPU--Wafer-Scale Accelerator Systems]{Characterizing Performance Tradeoffs in Heterogeneous GPU--Wafer-Scale Accelerator Systems}
\author{Ziyu Hu}
\email{zhu31@stevens.edu}
\affiliation{%
  \institution{Stevens Institute of Technology}
  \city{Hoboken}
  \state{New Jersey}
  \country{USA}
}

\author{Zhijing Ye}
\email{zye25@stevens.edu}
\affiliation{%
  \institution{Stevens Institute of Technology}
  \city{Hoboken}
  \state{New Jersey}
  \country{USA}
}

\author{Jiamin Wang}
\email{jwang259@stevens.edu}
\affiliation{%
  \institution{Stevens Institute of Technology}
  \city{Hoboken}
  \state{New Jersey}
  \country{USA}
}

\author{Zhiqing Zhong}
\email{zzhong9@stevens.edu}
\affiliation{%
  \institution{Stevens Institute of Technology}
  \city{Hoboken}
  \state{New Jersey}
  \country{USA}
}

\author{Zhendong Mi}
\email{zmi2@stevens.edu}
\affiliation{%
  \institution{Stevens Institute of Technology}
  \city{Hoboken}
  \state{New Jersey}
  \country{USA}
}

\author{Shaoyi Huang}
\email{shuang59@stevens.edu}
\affiliation{%
  \institution{Stevens Institute of Technology}
  \city{Hoboken}
  \state{New Jersey}
  \country{USA}
}

\author{Weijian Zheng}
\email{wzheng@anl.gov}
\affiliation{%
  \institution{Argonne National Laboratory}
  \city{Lemont}
  \state{Illinois}
  \country{USA}
}

\author{Rajkumar Kettimuthu}
\email{kettimut@anl.gov}
\affiliation{%
  \institution{Argonne National Laboratory}
  \city{Lemont}
  \state{Illinois}
  \country{USA}
}

\author{Xiaodong Yu}
\email{xyu38@stevens.edu}
\affiliation{%
  \institution{Stevens Institute of Technology}
  \city{Hoboken}
  \state{New Jersey}
  \country{USA}
}
\renewcommand{\shortauthors}{Hu et al.}

\begin{abstract}
As AI accelerator architectures become increasingly diverse, integrating different architectures into a unified system is emerging as a new heterogeneous computing paradigm. GPUs and WSEs differ substantially in compute, memory, and communication organization, creating both architectural complementarity and additional overheads, yet how to translate this heterogeneity into practical performance gains remains unclear. To address this gap, we propose a hierarchical benchmarking methodology to characterize GPU--WSE heterogeneous execution and the conditions for exploiting complementarity, together with an analytical performance model that identifies when heterogeneity becomes a bottleneck and when it yields system-level performance gains.

\end{abstract}

\maketitle

\section{Introduction}
GPUs have been remarkably successful in accelerating modern AI workloads due to their massive parallelism and mature software ecosystem, making them the de facto standard for both training and inference\cite{DBLP:conf/isca/LeeLBCJKKLIHKLR25,DBLP:journals/corr/abs-2511-11601}. However, as AI workloads become increasingly diverse, the limitations of general-purpose GPU architectures have become more apparent, driving the emergence of specialized accelerator architectures optimized for AI workloads. 

Wafer-scale accelerators represent an emerging class of specialized AI architectures that construct an entire wafer as a single computing device, rather than dividing it into multiple individual chips. For example, the Wafer-Scale Engine (WSE)\cite{lie2023cerebras} distributes approximately 900K compute cores and 44 GB of local SRAM across a single wafer-scale chip~\cite{alcf_cs3_system_overview}.

\underline{\textit{Motivation for GPU--WSE Integration:}} Modern AI workloads exhibit diverse computational and memory characteristics, making it difficult for a single accelerator architecture to efficiently serve all workloads. The complementary architectural characteristics of GPUs and WSEs provide an opportunity to address this challenge through heterogeneous integration. GPUs integrate a large number of highly parallel compute units and leverage SIMT execution to effectively utilize these resources, delivering high computational throughput. In contrast, WSEs leverage their massive chip area to integrate large amounts of distributed SRAM, allowing larger working sets to reside entirely on-chip and transforming much of the data movement that would otherwise rely on off-chip memory into on-chip accesses. Consequently, these distinct architectures exhibit different execution characteristics across AI workloads, creating opportunities for architectural complementarity in heterogeneous systems. This opportunity is also beginning to attract industrial interest, with emerging systems exploring the integration of AI accelerators with complementary architectural characteristics. For example, AWS and Cerebras have announced plans to integrate Trainium and WSE within a heterogeneous AI infrastructure\cite{cerebras_aws_collaboration_2026}.

\underline{\textit{Challenges in Exploiting GPU-WSE Complementarity:}} While the architectural differences between GPUs and WSEs create theoretical opportunities for complementarity, they also introduce substantial architectural asymmetry. Consequently, the feasibility boundaries of heterogeneous GPU--WSE execution have not been systematically studied. First, current AI system stacks are primarily designed for homogeneous accelerators or accelerators with similar architectures. Their parallelization strategies, system software, and interconnect mechanisms typically assume comparable hardware capabilities and execution characteristics, making them ill-suited to coordinating fundamentally different accelerators while fully exploiting their respective strengths.

Second, even though existing heterogeneous systems account for hardware differences, the forms of heterogeneity they address differ from those in GPU--WSE systems. For example, heterogeneous GPU systems~\cite{DBLP:journals/corr/abs-2404-14527,DBLP:journals/corr/abs-2406-01566,DBLP:journals/corr/abs-2507-10392,DBLP:journals/corr/abs-2605-04357,DBLP:conf/usenix/Kong0H25} primarily focus on heterogeneity across different generations of GPUs, while their underlying architectures remain largely the same. GPU--FPGA systems~\cite{DBLP:conf/ics/HeLXCCZL21,DBLP:journals/corr/abs-2102-01343} further introduce architectural heterogeneity, but FPGAs are general-purpose reconfigurable hardware whose design goals and execution characteristics differ substantially from those of specialized AI accelerators. Third, emerging heterogeneous platforms incorporating specialized AI architectures remain largely under development, and their performance characteristics remain largely unavailable to the research community. Consequently, existing heterogeneous system studies have not sufficiently explored heterogeneous cooperation between GPUs and specialized AI accelerators with substantially different architectures.

This leaves three fundamental questions: 1) Do the architectural differences between GPUs and WSEs create exploitable performance complementarity? 2) Under what application-level execution constraints can such complementarity translate into end-to-end performance gains? 3) What system support is required to realize these gains?

\underline{\textit{Our Approach:}} To answer these questions, we develop a hierarchical benchmarking methodology to systematically investigate the impact of architectural differences between GPUs and WSEs. At the application layer, we select three representative LLM execution paradigms---data parallelism, pipeline parallelism, and prefill--decode disaggregation---as case studies to analyze how architectural differences affect heterogeneous system performance. At the system layer, we investigate how compilers, communication mechanisms, and runtime systems should adapt to these architectural differences; at the hardware layer, we characterize the physical requirements for efficiently integrating these different architectures. \textbf{Observation:} 
\uline{Our results show that architectural difference is not inherently beneficial: its impact depends on whether the workload can be partitioned flexibly enough to accommodate architecture-specific asymmetry, and whether heterogeneous tasks can be mapped to the architectural strengths of different devices.} We further develop an analytical performance model that captures the key effects of architectural differences on heterogeneous execution. The main contributions are summarized as follows:
\begin{itemize}

\item We present a hierarchical, measurement-driven benchmarking methodology that examines how GPU--WSE architectural differences affect heterogeneous execution and its feasibility boundaries across the application, system, and hardware layers.

\item We apply this methodology to three representative LLM execution paradigms and reveal how different execution constraints can amplify, mitigate, or exploit GPU--WSE architectural differences. Based on these observations, we further identify the system and hardware support required to accommodate such architectural heterogeneity.

\item We develop a lightweight analytical model that formalizes these observations by capturing the interactions among workload decomposition, architectural asymmetry, resource utilization, and interconnect overhead. We validate the model against representative measured configurations and use it to characterize feasibility boundaries and reason about unmeasured GPU--WSE configurations.

\end{itemize}
\section{Background}
This section provides background on Wafer-Scale accelerators and commonly used parallelization strategies.
\subsection{Architectural Differences between GPUs and Wafer-Scale Accelerators}
\begin{figure}[ht]
   \begin{center}
   \includegraphics[width=\linewidth]{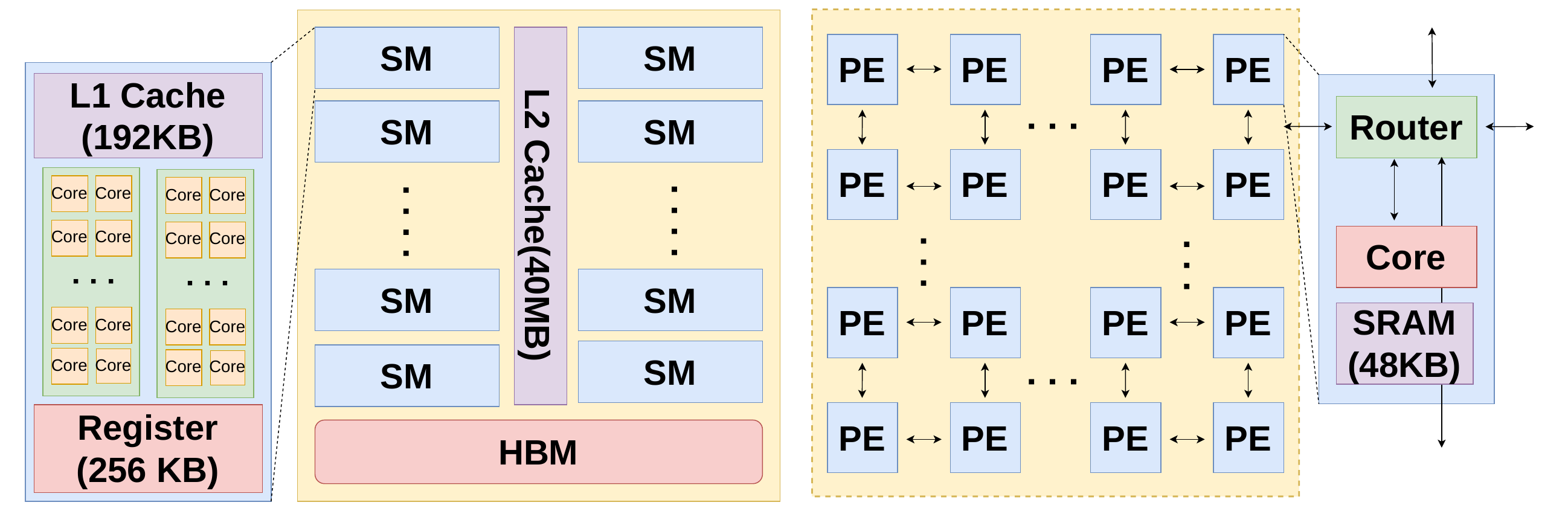}
   \caption{Architecture comparison between A100 (left) and WSE-3 (right), highlighting WSE's on-chip memory with higher capacity and closer proximity to each compute core compared to the GPU.}
   \vspace{-1em}
    \label{arch}
    \end{center}
\Description{A100 memory hierarchy and streaming multiprocessors on the left, and the wafer-scale processing-element mesh with local SRAM on the right.}
\end{figure}

GPUs and WSE exhibit significant architectural differences in the organization of compute, memory, and interconnection. We use the A100 and Cerebras WSE as representative architectures for comparison, as shown in Fig.~\ref{arch}.

\underline{\textit{Compute.}} The A100 consists of multiple Streaming Multiprocessors (SMs) and employs the SIMT execution model to schedule a large number of threads on each SM, effectively utilizing highly parallel compute units to achieve high computational throughput. In contrast, the WSE leverages the massive chip area enabled by wafer-scale integration to distribute approximately 900K processing elements (PEs) across a single wafer. Each PE contains an independent compute core, with computation partitioned and mapped across a large number of PEs and driven by dataflow execution. Thus, GPUs achieve high arithmetic throughput through densely integrated compute units and massive thread-level parallelism, whereas the WSE exploits wafer-scale integration to deploy fine-grained compute units for distributed execution across the wafer.

\underline{\textit{Memory.}} The A100 employs a hierarchical memory system consisting of on-chip registers, shared memory, and caches, together with off-chip HBM, providing multiple levels with different capacities and bandwidths. In contrast, the WSE leverages its wafer-scale area to integrate a large amount of distributed SRAM. Each PE is tightly coupled with up to 48KB of local SRAM, providing approximately 44GB of on-chip memory across the wafer. Thus, GPUs primarily rely on memory hierarchy and data reuse to reduce off-chip memory accesses, whereas the WSE allows larger working sets to be distributed and retained close to the compute cores, enabling more data accesses to remain on-chip.

\underline{\textit{Scaling.}} GPUs typically scale by adding discrete chips, with data exchange between GPUs relying on device-level interconnects such as NVLink. In contrast, the WSE can store model data in the external MemoryX system and stream it to the wafer on demand. For example, during training, the Weight Streaming mechanism sequentially transfers the weights of each layer from MemoryX to the wafer for execution. Thus, GPUs primarily scale computational capacity through multi-device parallelism, whereas a single WSE can execute the entire model computation.

\subsection{Cross-Device Execution Strategies for LLM Workloads}
Modern LLM systems rely on cross-device execution to distribute workloads and coordinate computation across multiple devices. During training, data parallelism (DP)~\cite{DBLP:conf/sc/RajbhandariRRH20} partitions the input data across devices, where each device maintains a model replica and synchronizes gradients through all-reduce. Pipeline parallelism (PP)~\cite{DBLP:journals/corr/abs-1811-06965} partitions the model into stages across devices and transfers intermediate activations between them. During inference, prefill--decode (PD) disaggregation~\cite{298687} places the prefill and decode stages on different devices, allowing the two stages to be executed and managed separately. We use these three representative cross-device execution paradigms as case studies to study architectural heterogeneity in GPU--WSE systems.
\section{Motivation}

This section presents the motivation for constructing heterogeneous systems that combine GPUs and wafer-scale architectures, as well as the key considerations involved.

\subsection{Motivation for GPU--WSE Heterogeneous Integration}

GPU--WSE heterogeneous systems provide an opportunity to address increasingly diverse AI workload demands through complementary hardware capabilities.

\textit{AI Workload Heterogeneity.} Modern AI workloads inherently consist of heterogeneous patterns. They present substantially different resource demands across compute, bandwidth, and capacity (Table~\ref{tab:workload_div}). Such diversity results in varying performance bottlenecks across AI workloads.

\begin{table}[t]{
  \centering

  \caption{Heterogeneous Workload Resource Requirements.}

  \label{tab:workload_div}
    \resizebox{1\linewidth}{!}{%
        \begin{tabular}{lccccc}
    \toprule
    Workload & Compute & Mem BW & Mem Cap  & Net BW & Storage Cap \\
    \midrule
    Prefill   & high   & low    & medium & medium      & low  \\
    Decode    & low    & high   & high   & low       & low  \\
    Training  & high   & high   & medium & high      & medium \\
    RAG       & low    & medium & high   & high       & high  \\
    Embedding & low    & high   & high   & low      & medium  \\
    \bottomrule
        \end{tabular}
    }
  }
\end{table}

\textit{Limitations of Homogeneous AI Systems.}
Platforms with homogeneous accelerators provide a fixed ratio of compute, bandwidth, and capacity per node, which is effective and scalable for workloads with consistent computation patterns. However, they become increasingly inefficient as workload heterogeneity grows, since different workloads are bottlenecked by different resource dimensions. As workload characteristics diverge, one resource configuration can no longer achieve efficient utilization, leading to a fundamental mismatch between workload requirements and system capabilities.

\textit{Hardware Heterogeneity as an Opportunity.} 
Different accelerator architectures exhibit distinct performance characteristics across resource dimensions (Table~\ref{tab:ai_acc}). This diversity enables complementary strengths across hardware platforms. By mapping workloads to the most suitable accelerators, heterogeneous systems can achieve better resource--demand alignment, improving both utilization and throughput.

\begin{table}[!htbp]
  \centering

  \caption{Heterogeneous Resource Provisioning.}

  \label{tab:ai_acc}
  \resizebox{1\linewidth}{!}{%
    \begin{tabular}{ccccccc}
      \toprule
      Vendor & Chip & Arch. & onchip & offchip & FLOPS & Power \\
      \midrule
      Cerebras & WSE3 & Wafer-Scale & 44\,GB & 13 TB & 125P & 23\,kW \\
      NVIDIA & A100 & Von Neumann & 0.19\,GB & 40\,GB & 312T & 400\,W \\
      \bottomrule
    \end{tabular}
  }

\end{table}

\subsection{Challenges in Exploiting GPU--WSE Architectural Complementarity}

As discussed above, while the architectural differences between GPUs and WSEs create potential complementarity, they also introduce substantial architectural asymmetry, which raises three key challenges:

\textbf{Realizing architectural complementarity through heterogeneous execution.}
Although GPUs and WSEs provide complementary hardware capabilities, whether such complementarity can be effectively exploited under heterogeneous execution remains unclear. Existing parallelization strategies are primarily designed for homogeneous systems and typically rely on uniform workload partitioning across devices. In heterogeneous GPU--WSE systems, hardware asymmetry breaks this assumption, while the tolerance of these strategies to performance asymmetry remains unclear. Moreover, whether these strategies can effectively exploit the architectural complementarity between GPUs and WSEs is also unclear.

\textbf{Adapting system support to architectural asymmetry.}
Realizing the potential benefits of GPU--WSE complementarity requires the system stack to coordinate devices with substantially different execution characteristics. Existing compiler, communication, and runtime mechanisms are primarily designed for homogeneous platforms and may not effectively accommodate such asymmetry. It therefore remains unclear what system support is required to efficiently coordinate heterogeneous execution.

\textbf{Meeting architectural requirements for heterogeneous platforms}.
Even with effective workload mapping and system support, the benefits of heterogeneous execution depend on whether the underlying hardware can efficiently support cross-device coordination. In particular, communication between architecturally distinct devices may offset their local performance advantages. It remains unclear what interconnect capabilities are required for GPU--WSE complementarity to translate into system-level performance gains.

\section{Hierarchical Benchmarking Design}
In this section, we present our hierarchical benchmarking study design and describe the benchmarking procedure for evaluating the GPU-WSE heterogeneous system across the three considerations.
\begin{figure}[ht]
   \begin{center}
   \vspace{-1em}
   \includegraphics[width=0.95\linewidth]{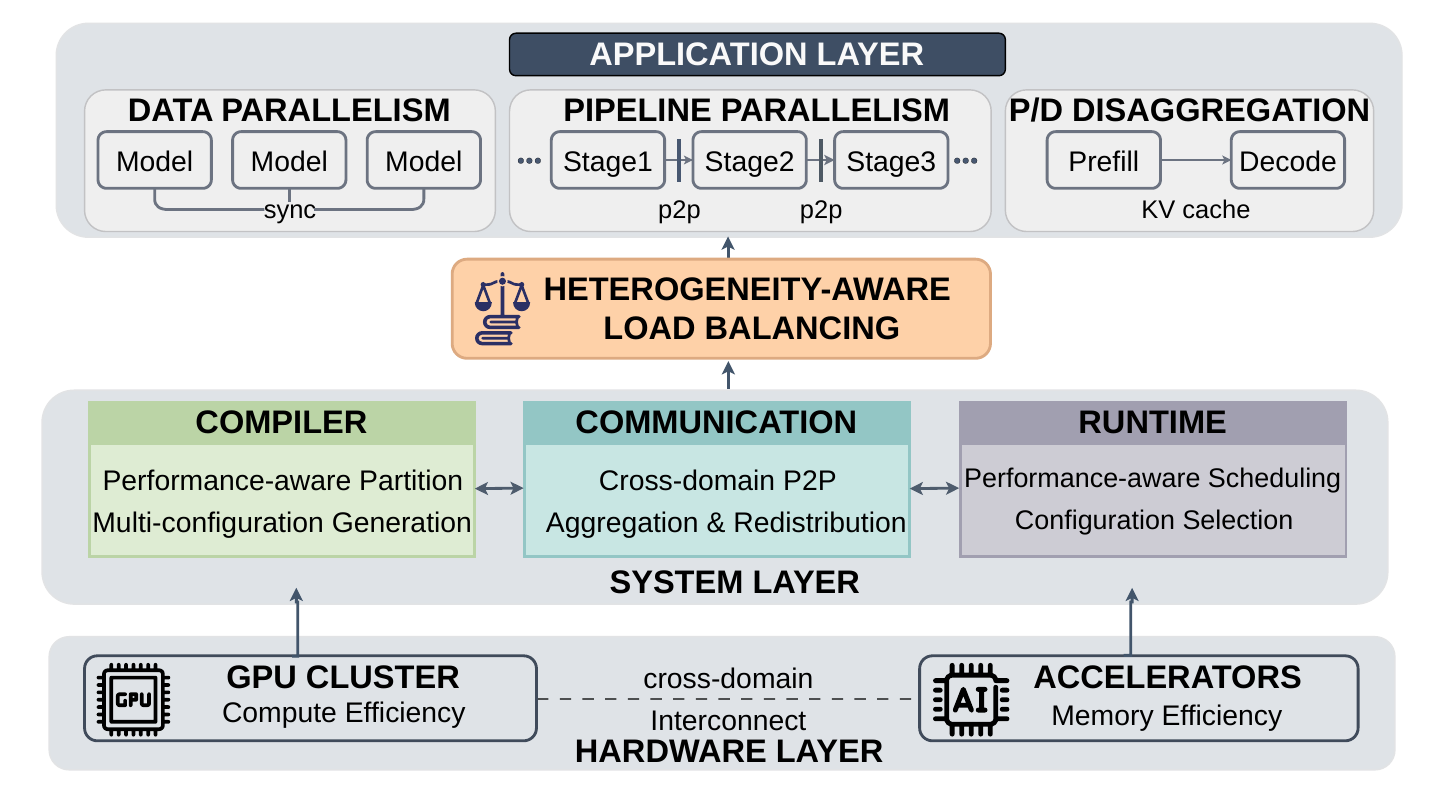}
   \caption{Overview of hierarchical benchmarking design for GPU-WSE heterogeneous systems. The benchmarking methodology across three layers: application, system, and hardware, corresponding to the three considerations.}
   \vspace{-1.5em}
  \label{framework}
    \end{center}
\Description{Three-layer benchmarking framework linking application execution strategies, compiler and communication support, runtime scheduling, and hardware interconnects.}
\end{figure}

\subsection{Benchmarking Framework}
To address the three challenges discussed above, we develop a three-tier benchmarking framework spanning the \textit{application}, \textit{system}, and \textit{hardware} layers, as illustrated in Fig.~\ref{framework}. The framework systematically explores how GPU--WSE architectural differences affect heterogeneous execution: the application layer investigates what characteristics of parallelization strategies determine their tolerance to architectural asymmetry and ability to exploit architectural complementarity; the system layer investigates how compiler, communication, and runtime mechanisms should adapt to architectural asymmetry; and the hardware layer examines the interconnect conditions under which the potential benefits of heterogeneous execution can translate into end-to-end performance gains. We describe each layer in detail as follows:

\textit{Application Layer.}
At this layer, we investigate what execution characteristics enable parallelization strategies to tolerate GPU--WSE performance asymmetry and exploit the complementarity between the two architectures. For heterogeneous execution, a key distinction lies in the constraints imposed by different execution paradigms. Based on existing execution models, we focus on two representative coordination paradigms: BSP-style execution follows a structured computation--communication--synchronization pattern with explicit global coordination, whereas task-based execution organizes computation as dependency-driven tasks, enabling more flexible task scheduling and resource mapping. These two paradigms represent different degrees and forms of execution constraints, providing representative contrasts for analyzing heterogeneous execution.

\textit{System Layer.}
At this layer, we investigate how the existing system stack should adapt to GPU--WSE architectural asymmetry to support efficient heterogeneous execution. Such architectural asymmetry introduces three key requirements for the system stack: generating asymmetric workload mappings that account for distinct hardware capabilities, supporting asymmetric communication across architectures with different communication characteristics, and dynamically coordinating execution under performance asymmetry across devices. These requirements correspond to the core functionalities of the compiler, communication mechanisms, and runtime, respectively. We therefore analyze the system support required for GPU--WSE heterogeneous execution from these three aspects.

\textit{Hardware Layer.}
At this layer, we examine whether the potential benefits of heterogeneous execution can translate into end-to-end performance gains under interconnect constraints. Heterogeneous execution inevitably introduces cross-architecture communication, and insufficient interconnect capability may amplify the impact of architectural asymmetry or offset architectural complementarity, potentially making heterogeneous execution less efficient than single-architecture execution. We therefore characterize the minimum interconnect bandwidth required for effective GPU--WSE heterogeneous execution.

\subsection{Benchmarking Procedure}
\begin{figure}[ht]
   \begin{center}
   \vspace{-1em}
   \includegraphics[width=0.9\linewidth]{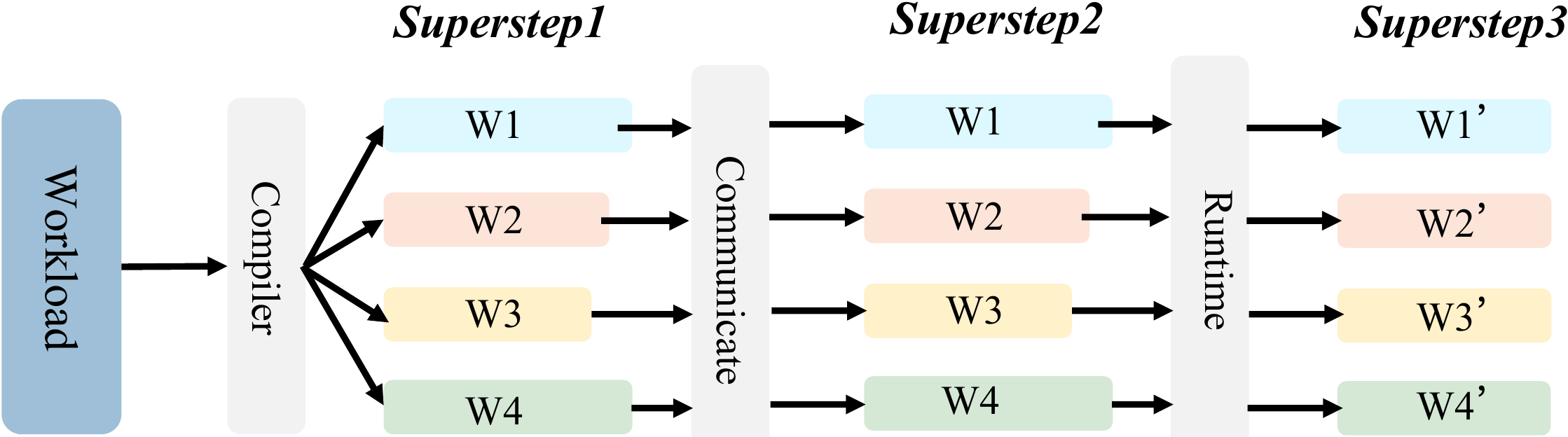}
   \caption{BSP workflow across heterogeneous hardware. Different colors represent different devices, and the length of each block denotes the execution time within a superstep. }

  \label{bsp}
    \end{center}
\Description{Three synchronized supersteps with computation, communication, and runtime coordination across four device workloads.}
\end{figure}

\textit{BSP-Style Execution.} As illustrated in Fig.~\ref{bsp}, BSP-style execution organizes computation into a sequence of supersteps, with a global synchronization barrier at the end of each superstep. When the GPU and WSE operate at different execution rates, the faster domain must wait for the slower one. Workload partitioning can mitigate this imbalance by adjusting the amount of work assigned to each domain, but its effectiveness is constrained by the partitioning dimensions and execution constraints of each parallelization strategy.
\begin{figure}[t]
    \centering
    \includegraphics[width=0.65\linewidth]{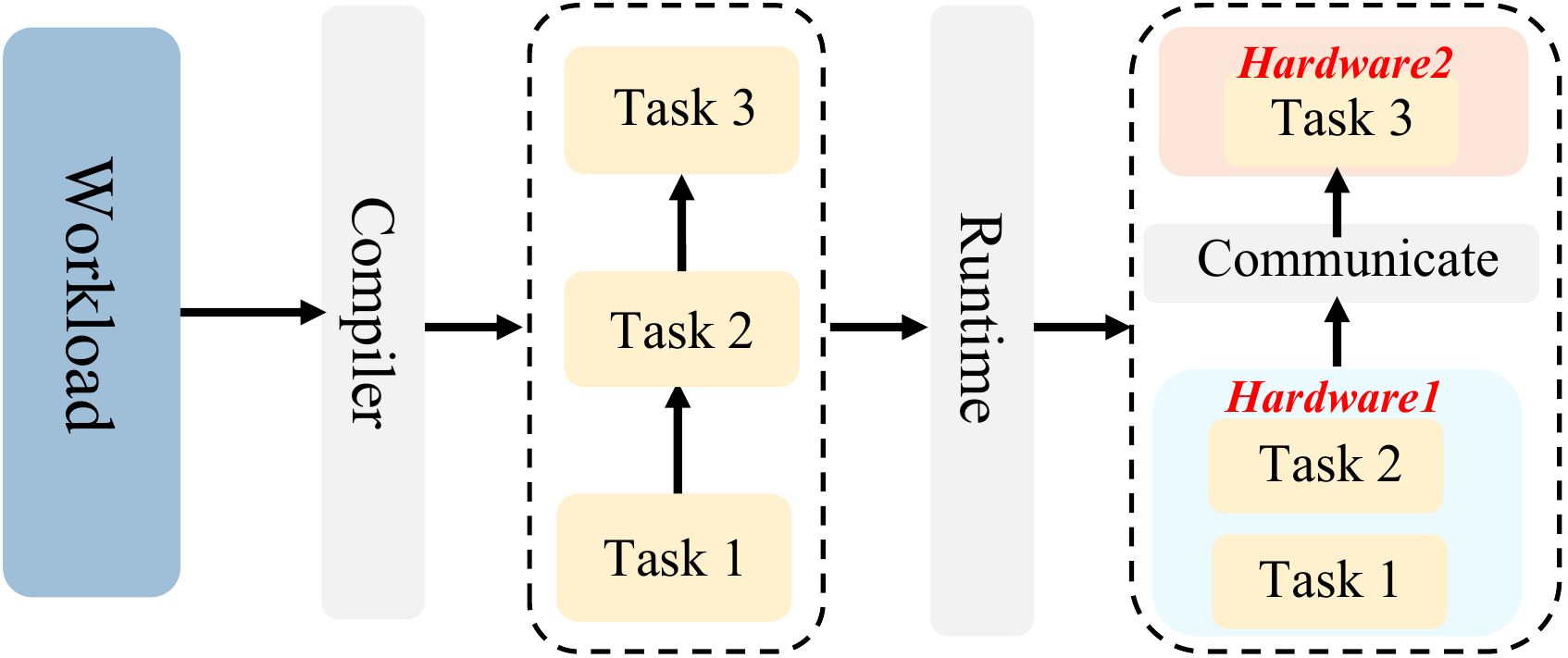}
    \caption{Task-style workflow across heterogeneous hardware.
    The workload is decomposed into a task DAG and
    scheduled across heterogeneous devices.}
    \Description{A task-based workflow illustrating how
    workload tasks are scheduled across heterogeneous devices.}
    \label{fig:task_based}
\end{figure}

\begin{figure*}[t]
    \centering
    \includegraphics[width=0.98\linewidth]{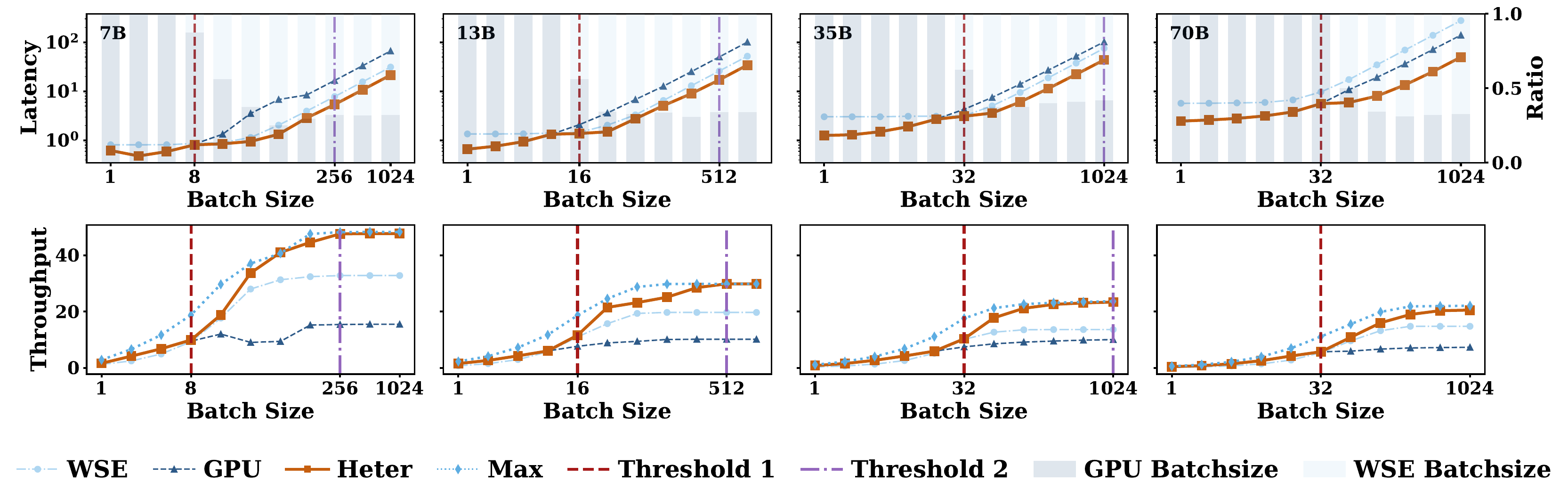}

    \caption{Step latency (s) and throughput (samples/s) under varying
    global batch sizes (left to right: 7B, 13B, 35B, and 70B).
    WSE, GPU, Heter, and Max identify
    WSE-only, GPU-only, WSE--GPU hybrid data parallelism
    (hybrid DP), and the aggregate standalone throughput bound, respectively.
    Threshold 1 and Threshold 2 mark the points where hybrid
    DP starts to yield performance gains and where the gains
    are fully saturated, respectively. The bars indicate the
    batch size allocated to each platform.}

    \Description{Performance comparison of WSE-only,
    GPU-only, and hybrid data parallelism across four
    model sizes and varying global batch sizes.}

    \label{fig:dp_latency}
\end{figure*}

We select DP and PP as two representative cases of BSP-style execution to study how different coordination constraints shape the impact of architectural differences. DP is primarily constrained by BSP-style global synchronization, while batch-dimension partitioning provides relatively high flexibility and loose coordination between the GPU and WSE. In contrast, PP introduces additional pipeline structure, stage partitioning, and finer-grained cross-hardware coordination constraints. By comparing these cases from coarse- to fine-grained coordination, we examine how increasing execution constraints change the impact of GPU--WSE architectural differences on heterogeneous execution.

\textit{Task-Based Execution.} As illustrated in Fig.~\ref{fig:task_based}, task-based execution decomposes the workload into dependency-driven tasks and allows them to be independently mapped onto heterogeneous hardware. Unlike BSP-style execution, it does not require the two hardware domains to maintain matched execution progress at global synchronization boundaries, but is primarily constrained by task dependencies and cross-domain data transfers. Unlike DP and PP, which mainly partition homogeneous computation in different ways, we select PD disaggregation as a representative case because prefill and decode naturally form two heterogeneous execution stages with distinct compute and memory characteristics, allowing us to study the impact of heterogeneity within the workload on heterogeneous execution.

\section{Performance Evaluation and Analysis}
In this section, we present experimental results and analysis for three parallelization strategies in heterogeneous systems. We further summarize the key insights of different execution paradigms in heterogeneous settings.

\subsection{Experimental Setup}

We evaluate GPU--WSE heterogeneous execution using computing resources at the Argonne Leadership Computing Facility (ALCF), including measurements from a Cerebras WSE-3 cluster and an NVIDIA DGX A100 GPU cluster. We consider models with 7B, 13B, 35B, and 70B parameters, and select hardware configurations for each experiment according to the deployment and parallelization requirements of the corresponding workload. DP and PP are evaluated under training workloads, while PD disaggregation is evaluated under inference workloads. We independently measure the execution characteristics of the GPU and WSE on real hardware and combine them using the calibrated discrete-event simulator SimAI~\cite{DBLP:conf/nsdi/WangLXL0CZZZZLZ25}. SimAI is designed for large-scale GPU-cluster simulation and supports user-defined workload and measured computation-time inputs, allowing us to incorporate the measured WSE behavior into heterogeneous execution simulation. By replaying measured configurations, we verify that the simulated results closely match the corresponding hardware measurements.

\subsection{BSP-Style Heterogeneous Data Parallelism}
\subsubsection{Overall Performance Results}

Figure~\ref{fig:dp_latency} shows the step latency and throughput of the GPU, WSE, and heterogeneous configurations across different global batch sizes. The latency results additionally show the GPU--WSE batch partition that minimizes the heterogeneous step latency, while the throughput results include the theoretical aggregate upper bound obtained by summing the standalone throughput of the two hardware domains.

The results reveal three distinct execution regimes separated by two workload thresholds. At small workloads, architectural differences primarily manifest as \textit{latency asymmetry}. Neither architecture is sufficiently utilized, while their different architectures exhibit different baseline execution latencies. Under BSP synchronization, assigning part of the workload to the higher-latency domain increases the synchronized step time, making single-domain execution preferable. The first threshold therefore marks the point beyond which heterogeneous workload partitioning begins to compensate for this latency asymmetry.

\begin{figure}[htbp]
    \centering
    
    \begin{minipage}[t]{\linewidth}
        \centering
        \includegraphics[width=\linewidth,height=0.09\textheight]{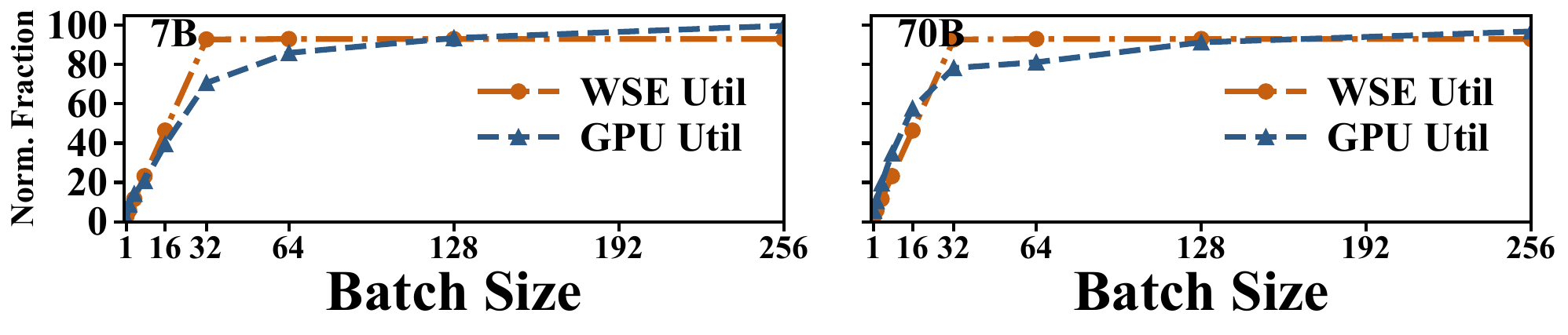}
        \par

        \centering
    \end{minipage}%

    \caption{Utilization scaling asymmetry. GPU utilization is measured using MFU following~\cite{DBLP:journals/jmlr/ChowdheryNDBMRBCSGSSTMRBTSPRDHPBAI23}. WSE utilization is computed as the product of PE allocation ratio and the utilization of the allocated PEs following~\cite{DBLP:conf/iiswc/HuZZYTZXKY25}. We compare utilization trends within each platform, not absolute values.}
    
    \label{fig:dp_para}

\Description{GPU and WSE utilization curves across batch sizes for the 7B and 70B models.}
\end{figure}

Beyond the first threshold, the system enters a \textit{scaling-asymmetry regime}. As shown in Fig.~\ref{fig:dp_para}, the two architectures exhibit substantially different utilization-scaling behaviors as the batch size increases. This difference stems from their distinct execution organizations and compute scales: the GPU domain can achieve high compute-resource utilization under relatively small workloads, whereas the wafer-scale architecture requires larger workloads to expose sufficient parallelism across its large PE array. In our measurements, the WSE does not achieve high utilization until the batch size approaches 32. These distinct utilization-scaling behaviors further result in different performance-scaling curves, causing the relative processing capabilities of the GPU and WSE to vary with workload size. Consequently, the optimal GPU--WSE batch partition also changes substantially throughout this regime.

As the workload further increases, both architectures gradually enter \textit{stable utilization regime}, forming the second threshold. Beyond this point, their relative processing capabilities stabilize, the optimal workload partition converges, and heterogeneous throughput approaches the theoretical aggregate upper bound. At this stage, architectural asymmetry is largely absorbed through workload partitioning.

\subsubsection{Layer-Wise Performance Analysis}

\paragraph*{Application Layer Analysis}
The three execution regimes show that, architectural differences primarily manifest as architectural asymmetry rather than exploitable architectural complementarity for DP. BSP synchronization exposes this asymmetry, while flexible partitioning along the batch dimension provides a degree of freedom to absorb it as workload increases. However, heterogeneous throughput remains bounded by the aggregate standalone capacity of the GPU and WSE, indicating that workload partitioning mitigates architectural asymmetry but does not convert architectural differences into additional complementarity.

\paragraph*{System Layer Analysis.}
Although DP cannot exploit architectural complementarity to provide additional performance gains, proper workload partitioning can still enable heterogeneous execution to approach the aggregate capacity of the GPU and WSE, thereby utilizing otherwise idle heterogeneous hardware resources. To achieve this, the system stack must effectively accommodate architectural asymmetry. First, the compiler needs to generate asymmetric workload partitions according to the different execution capabilities of the two architectures. Second, in the scaling-asymmetry regime, the relative performance of the GPU and WSE varies with workload scale, making the optimal batch partition unstable. The runtime therefore needs to dynamically adjust workload allocation based on observed execution behavior to maintain balanced execution across the two hardware domains. Finally, because large models are typically distributed across multiple devices in the GPU domain but organized within a single wafer-scale execution domain on the WSE, the two domains use different distributed representations. Cross-domain synchronization therefore requires the communication layer to perform the necessary gradient aggregation and redistribution to ensure correct global synchronization.
\begin{figure}[ht]
   
   \begin{center}
   \includegraphics[width=0.8\linewidth]{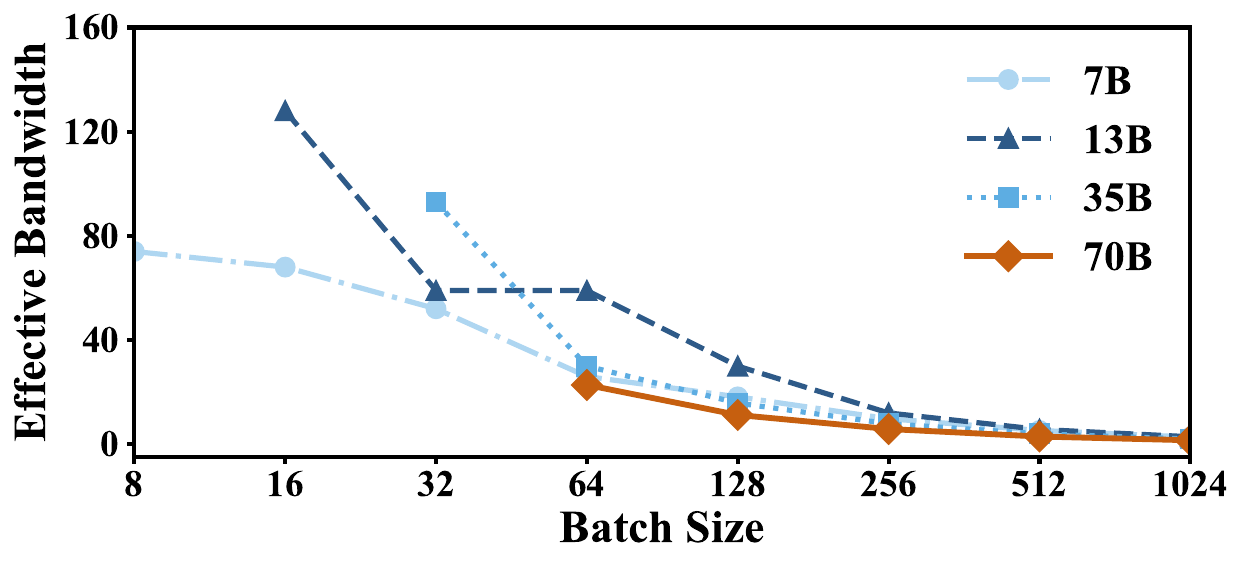}

   \caption{Effective bandwidth (GB/s) lower bound for heterogeneous execution to achieve throughput gains under varying global batch sizes.}
  \label{dp_bw}
    \end{center}

\Description{Effective interconnect bandwidth thresholds across global batch sizes for heterogeneous data parallel execution.}
\end{figure}

\paragraph*{Hardware Layer Analysis}
To enable heterogeneous DP to provide performance benefits, we further characterize the lower bound on the interconnect bandwidth required to preserve the gains from heterogeneous collaboration, as shown in Fig.~\ref{dp_bw}. Below this lower bound, the benefits of heterogeneous execution are entirely offset by cross-domain communication overhead. As workload size increases, the required bandwidth lower bound gradually decreases.

\begin{tcolorbox}[colback=gray!20, 
  colframe=black, 
  boxsep=2pt,        %
  left=2pt,          %
  right=2pt,         %
  top=2pt,           %
  bottom=2pt         %
  ]
\textbf{Takeaway 1}:  GPU-WSE architectural differences are primarily performance obstacles in DP. However, DP's loose coupling allows flexible system-level workload partitioning to effectively absorb this asymmetry at sufficiently large workloads.
\end{tcolorbox}

\subsection{BSP-Style Heterogeneous Pipeline Parallelism}

\subsubsection{Overall Performance Results}

\paragraph*{Standard PP}

\begin{figure}[ht]

   \begin{center}
   \includegraphics[width=0.8\linewidth]{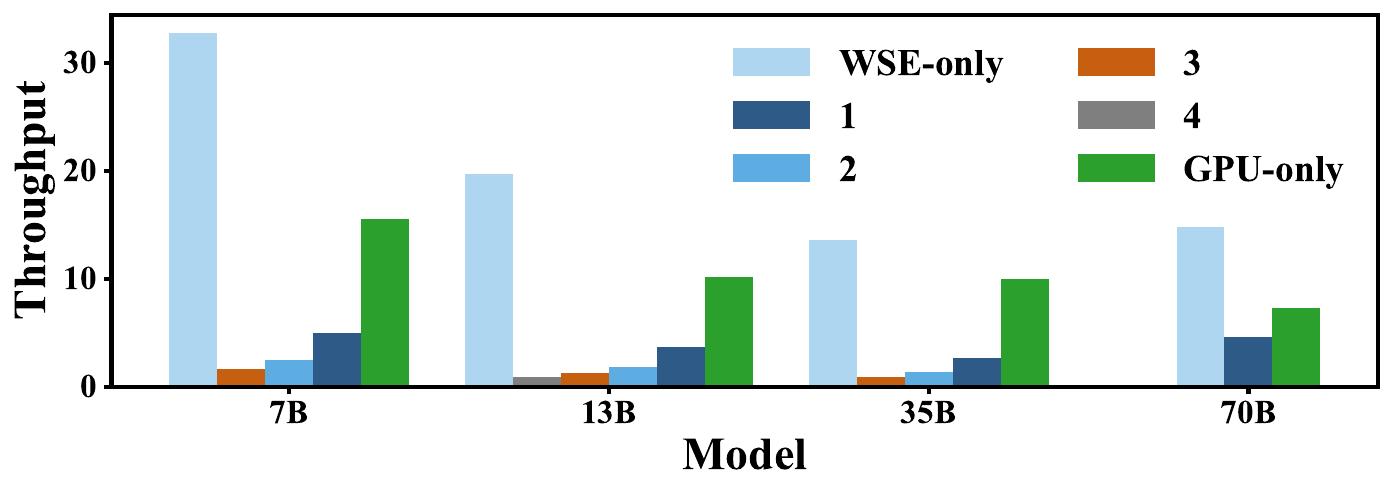}

   \caption{Throughput at global batch size 1024 and micro-batch size 1 under GPU, WSE, and hybrid PP placements. Labels denote the number of layers offloaded from each GPU stage to WSE, relative to GPU baseline. Offloading is uniform across stages because PP is constrained by the slowest stage, and partial offloading does not alter bottleneck.}
  \label{fig:micro_1}
    \end{center}

\Description{Throughput comparison of standalone and heterogeneous pipeline placements at micro-batch size one.}
\end{figure}

Figure \ref{fig:micro_1} shows the results of heterogeneous pipeline parallelism (PP) across four model sizes with a micro-batch size of 1, including GPU-only, Cerebras-only, and a range of heterogeneous configurations, where a fixed number of layers are offloaded from each GPU stage to the WSE while the remaining layers are retained on the GPU. In terms of throughput, none of the heterogeneous PP configurations outperform the single-platform setups.

\begin{figure}[htbp]
    \centering
    
    \begin{minipage}[t]{\linewidth}
        \centering
        \includegraphics[width=\linewidth,height=0.11\textheight]{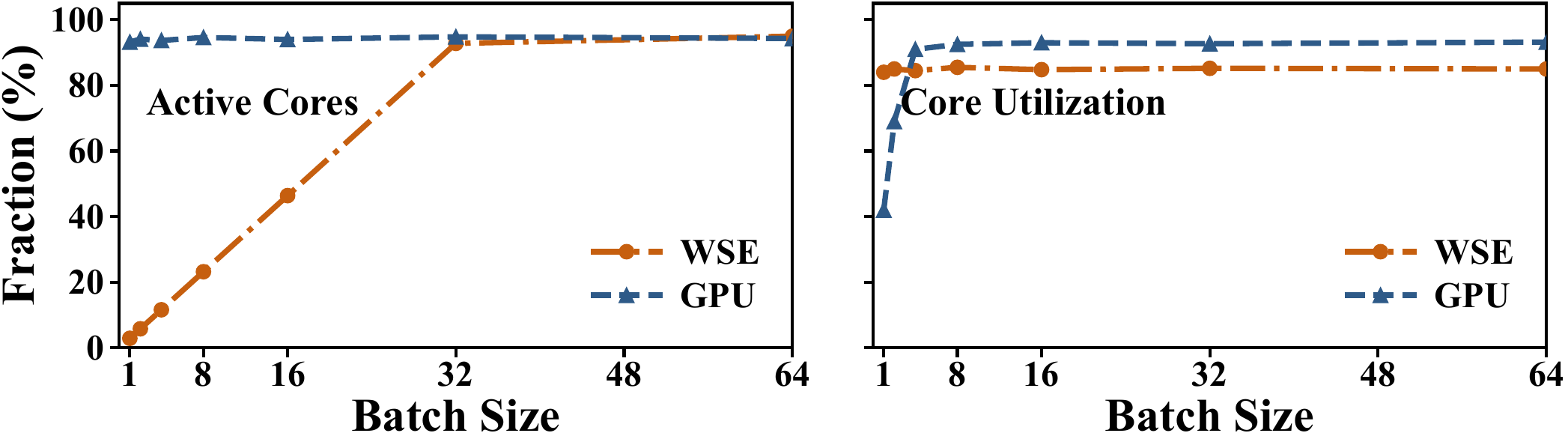}
        \par

        \centering
    \end{minipage}%

    \caption{Compute-core activation rate and utilization. For GPUs, compute cores refer to streaming processors, while for WSEs they refer to processing elements (PEs).}
    
    \label{fig:core}

\Description{Compute-core activation and active-core utilization across batch sizes for GPUs and WSEs.}
\end{figure}

As shown in Fig.~\ref{fig:core}, under LLM workloads, the utilization of active cores remains relatively stable on both platforms, indicating that overall utilization is primarily determined by the fraction of activated cores. GPUs can activate a large fraction of their SMs even at very small batch sizes, whereas the WSE requires a batch size of at least 32 to activate nearly all of its cores due to its wafer-scale architecture. Under PP, however, all hardware domains must use a unified micro-batch size, preventing the WSE from operating in its preferred large-batch regime and thereby limiting its active-core fraction. Consequently, when one layer from each GPU stage is offloaded to the WSE, the WSE processes these layers under low utilization and eventually becomes the bottleneck of the entire pipeline.

\begin{figure*}[ht]
   \begin{center}

   \includegraphics[width=\linewidth,height=0.15\textheight]{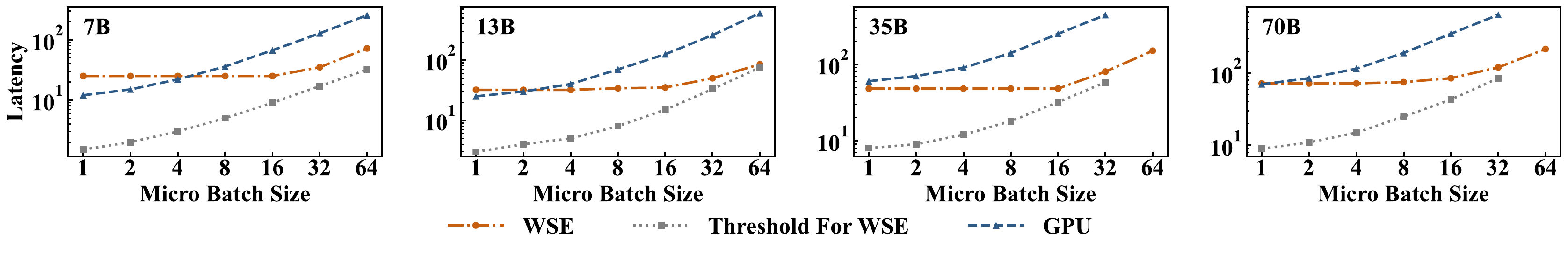}

   \caption{Per-layer execution latency(ms) of GPU and WSE across different micro-batch sizes, and the thresholds (vs. GPU-only) at which offloading one layer per GPU stage to WSE yields performance benefits.}
  \label{fig:pp_latency}
    \end{center}

\Description{Per-layer GPU and WSE execution latency versus micro-batch size, with the heterogeneous pipeline break-even threshold.}
\end{figure*}
To determine whether any heterogeneous PP configuration can yield performance gains, we analyze the per-layer latency of the GPU and WSE under different micro-batch sizes. As shown in Fig.~\ref{fig:pp_latency}, among all heterogeneous PP configurations, offloading one layer from each GPU stage to the WSE achieves the highest throughput. We therefore use this configuration as a representative case and define a latency threshold for the WSE, as shown in Fig.~\ref{fig:pp_latency}. This threshold represents the critical execution latency that the WSE must achieve when processing the offloaded layers in order to avoid becoming the slowest stage in the pipeline.
\begin{figure}[htbp]

    \centering
    \begin{minipage}[t]{\linewidth}
        \centering
        \includegraphics[width=\linewidth]{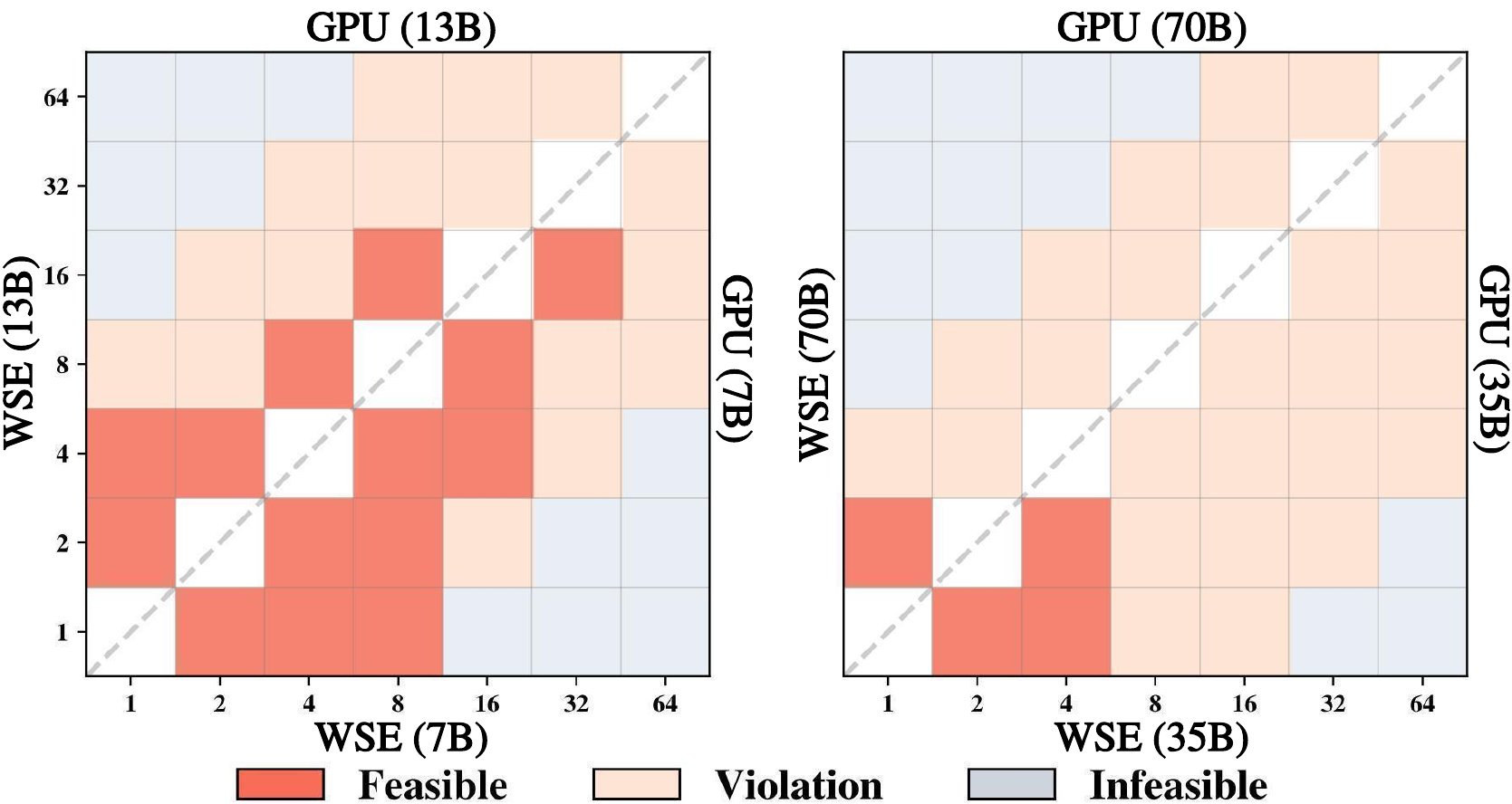}

    \end{minipage}\hfill
           \caption{Feasible placements under varying micro-batch sizes. Infeasible denotes placements that violate pipeline structural constraints; violation denotes placements exceed memory constraints; feasible denotes valid placements.}
\label{fig:feas}
\Description{Heatmaps indicating feasible, structurally infeasible, and memory-violating heterogeneous pipeline placements for four model sizes.}
\end{figure}
As the micro-batch size increases, the per-layer latency of the WSE gradually approaches this threshold. This indicates that, as WSE utilization improves, the performance gap between the WSE and GPU continues to narrow, making heterogeneous execution increasingly likely to provide performance benefits. Although the two curves do not intersect under our current hardware configuration, such an intersection may occur under a different GPU/WSE performance balance. However, due to the memory asymmetry, GPUs cannot in practice support such large micro-batch sizes, preventing the WSE from entering its high-utilization regime.

\paragraph*{Optimized PP}

The previous results show that architectural differences between GPUs and WSEs lead to different preferred workload scales. Forcing different architectures to process workloads of the same scale can prevent each platform from fully exploiting its architectural advantages.

Based on this observation, we further explore an architecture-aware heterogeneous PP scheme that allows GPUs and the WSE to use different micro-batch sizes. Multiple GPU pipelines are paired with a single WSE pipeline, enabling the WSE to aggregate smaller GPU micro-batches into a larger workload. As shown in Fig.~\ref{fig:feas}, feasible configurations exist for all models, demonstrating that workload-scale decoupling can partially accommodate the different preferred workload scales of the two architectures. However, because PP requires fine-grained coordination, memory asymmetry and pipeline structural constraints still limit the freedom of the two architectures to independently select their workload scales, thereby substantially reducing placement flexibility.
\begin{figure}[ht]
   \begin{center}

   \includegraphics[width=0.8\linewidth]{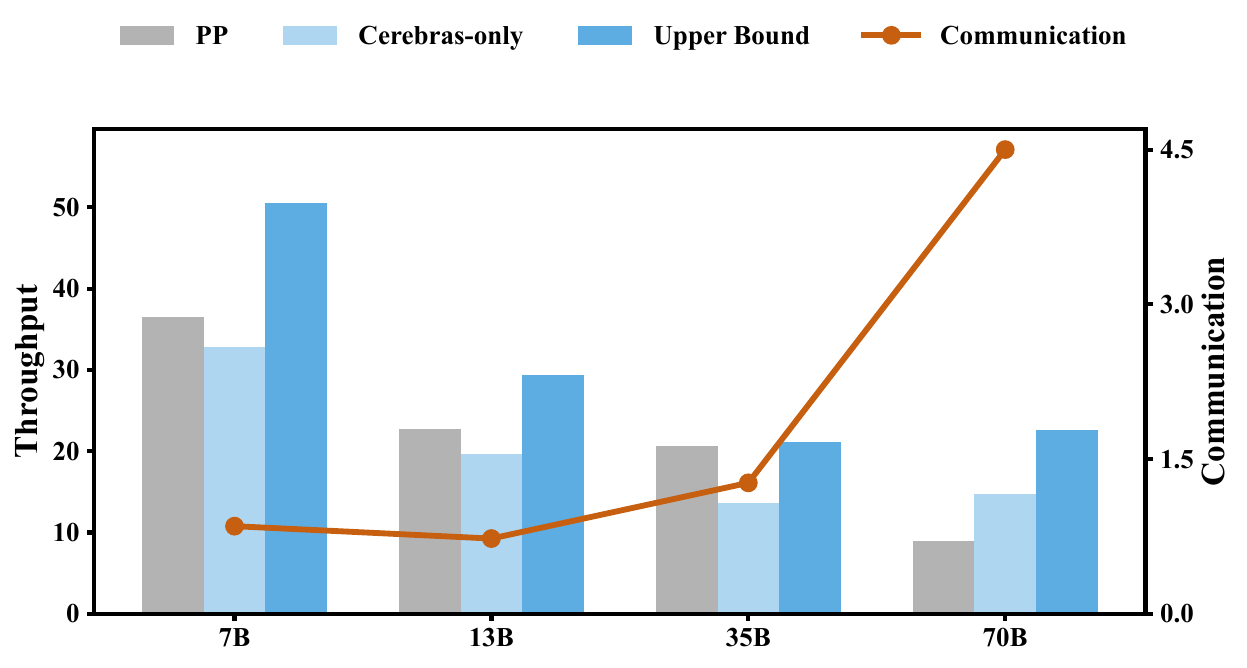}
   \caption{Throughput (sample/s) of optimized PP vs. WSE-only, and network bandwidth requirements.}

  \label{fig:pp_result}
    \end{center}
\Description{Optimized pipeline throughput relative to the WSE-only baseline and associated interconnect bandwidth requirements.}
\end{figure}

\begin{figure*}[htbp]
    \centering

    \begin{minipage}[t]{\linewidth}
        \centering
        \includegraphics[width=\linewidth]{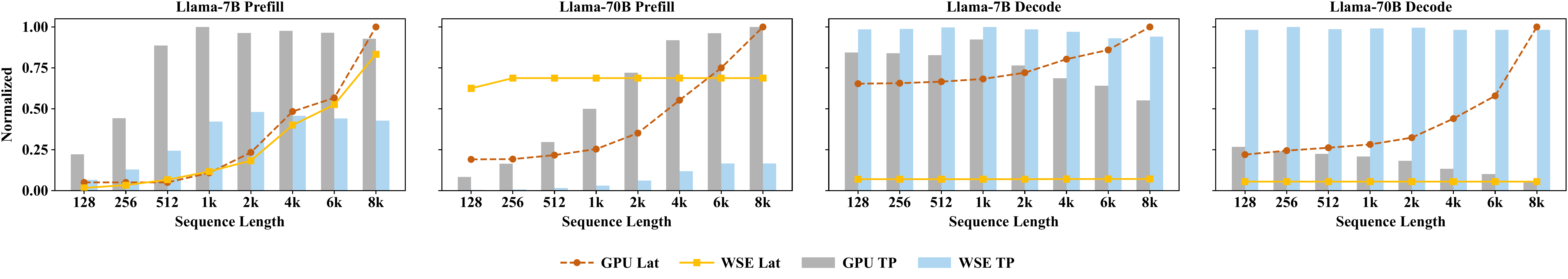}
        \par

        \centering
    \end{minipage}%

    \caption{Normalized prefill and decode performance across sequence lengths. Lines denote latency and bars denote throughput. From left to right: 7B prefill, 70B prefill, 7B decode, and 70B decode.}

    \label{fig:prefill}
\Description{Four panels show normalized latency and throughput by sequence length for 7B prefill, 70B prefill, 7B decode, and 70B decode.}
\end{figure*}

Finally, for each model, we select a feasible configuration that maintains relatively high WSE utilization and compare its throughput at a global batch size of 1024 against the best-performing WSE-only baseline in Fig.~\ref{fig:pp_result}. From the figure, the 7B, 13B, and 35B models all achieve higher throughput. However, these configurations still do not reach the theoretical performance upper bound. This indicates that although workload-scale decoupling can alleviate the utilization mismatch caused by architectural differences, the structural constraints of PP and memory asymmetry still limit the system's ability to absorb such heterogeneity, preventing the potential complementarity between the two architectures from being fully exploited.

\subsubsection{Layer-Wise Performance Analysis}

\paragraph*{Application Layer Analysis}
The two PP experiments show that, under BSP-style PP execution, architectural differences between GPUs and WSEs likewise primarily manifest as architectural asymmetry rather than exploitable architectural complementarity. Unlike DP, which can flexibly adjust the workload assigned to different hardware domains along the data dimension and use uneven data partitioning to absorb performance asymmetry as much as possible, PP partitions the workload along the layer dimension and inherently provides less workload-partitioning freedom. This freedom is further restricted by constraints such as a unified micro-batch size, GPU memory capacity, and pipeline structure. As a result, PP has limited ability to flexibly match workloads to the performance characteristics of different architectures, making architectural asymmetry difficult to fully absorb and more likely to translate directly into a pipeline bottleneck.

\paragraph*{System Layer Analysis}
PP is not an ideal parallelization strategy for GPU--WSE heterogeneous systems, but it can still provide performance benefits under certain configurations by utilizing otherwise idle heterogeneous hardware resources. Therefore, if heterogeneous PP is considered, the most important first step is to determine whether it is actually beneficial. At the compiler level, architectural asymmetry may cause heterogeneous execution to perform worse than execution on a single hardware domain. The compiler therefore needs to perform a feasibility check to determine whether heterogeneous PP can provide performance gains under a given configuration or whether execution should fall back to a single hardware domain. At the same time, PP has a large configuration search space, making static compiler decisions prone to misprediction. The runtime therefore needs to correct suboptimal compiler decisions based on observed execution behavior and trigger fallback or reconfiguration when necessary. At the communication layer, in addition to providing cross-domain P2P transfers, the system must, as in DP, handle inconsistencies in distributed activation representations across different hardware domains.

\paragraph*{Hardware Layer Analysis} For heterogeneous PP, the interconnect still needs to provide sufficient bandwidth for basic cross-domain communication. Using the heterogeneous PP configurations in Fig.~\ref{fig:pp_result} as representative cases, we report the effective communication bandwidth required to preserve their performance benefits. Depending on the model size and configuration, the required bandwidth ranges from several hundred MB/s to a few GB/s, providing a practical reference range for GPU--WSE interconnect design.

\begin{tcolorbox}[colback=gray!20, 
  colframe=black, 
  boxsep=2pt,        %
  left=2pt,          %
  right=2pt,         %
  top=2pt,           %
  bottom=2pt         %
  ]
\textbf{Takeaway 2}:  Under PP, GPU--WSE architectural differences primarily manifest as performance obstacles. PP's fine-grained coordination and multiple execution constraints amplify the negative effects of architectural asymmetry. As a result, even with substantial system-level effort to accommodate these differences, only limited performance gains can be achieved.
\end{tcolorbox}

\subsection{Task-Based Heterogeneous PD Disaggregation}

\subsubsection{Overall Performance Results}
\leavevmode\par
Figure~\ref{fig:prefill} shows the performance of the two hardware platforms under different request shapes. In the \textbf{prefill stage}, GPU throughput increases rapidly with sequence length and consistently remains higher than that of the WSE. In contrast, during the \textbf{decode stage}, the WSE achieves significantly higher throughput and better latency scalability: as the context length increases, GPU decode latency grows from tens to hundreds of milliseconds, whereas the WSE maintains relatively stable microsecond-level latency across different settings.
\begin{figure}[ht]
   \begin{center}

   \includegraphics[width=0.8\linewidth]{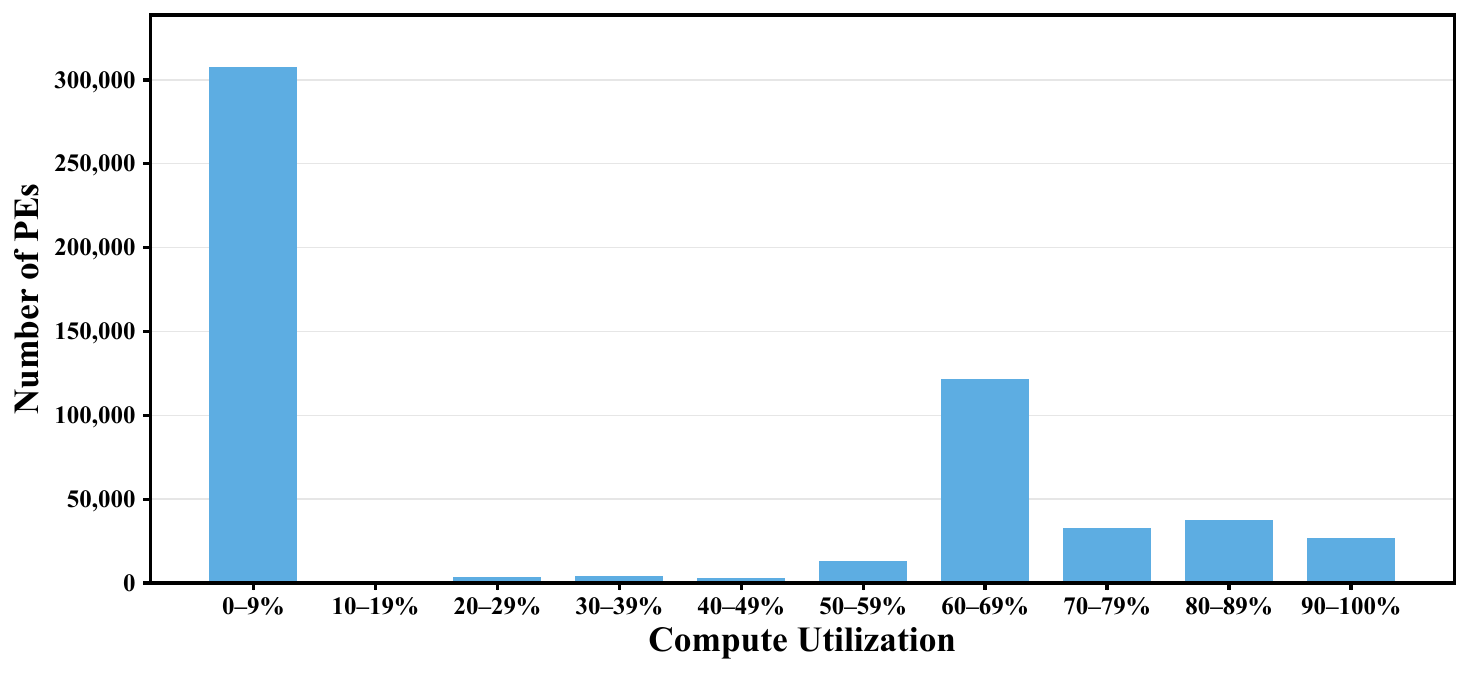}
   \caption{WSE PE Utilization Snapshot. Only a small fraction of compute cores achieve utilization above 90\%.}

  \label{fig:pe_u}
    \end{center}
\Description{Distribution of WSE processing-element utilization during a representative execution interval.}
\end{figure}

 GPUs achieve better performance during prefill for two main reasons. First, the GPU's SIMT execution model allows a large number of cores to execute the same instructions on different data elements, making it well suited to regular, dense matrix computations. In contrast, the WSE's dataflow execution model spatially maps different kernels onto different PE regions, which can produce load imbalance and cause overall performance to be limited by the most heavily loaded region. Second, GPUs operate at a smaller compute scale. Large matrix operations are typically divided into multiple tiles and processed in successive waves, allowing even relatively small workloads to maintain high core utilization. The WSE, by comparison, operates at wafer scale and requires a larger workload to fully utilize its on-chip PEs. Moreover, the shape of a matrix operation may not map efficiently onto its large-scale PE array, leaving some compute resources underutilized. To validate this analysis, we examined PE utilization during a representative interval of WSE execution. As shown in Fig.~\ref{fig:pe_u}, a large fraction of the PEs exhibit low compute utilization, while only a small fraction achieve utilization above 90\%. These results show that the WSE's execution model and wafer-scale architecture make it more challenging to fully utilize its on-chip compute resources.

\begin{figure}[htbp]

    \centering
    \begin{minipage}[t]{\linewidth}
        \centering
        \includegraphics[width=\linewidth]{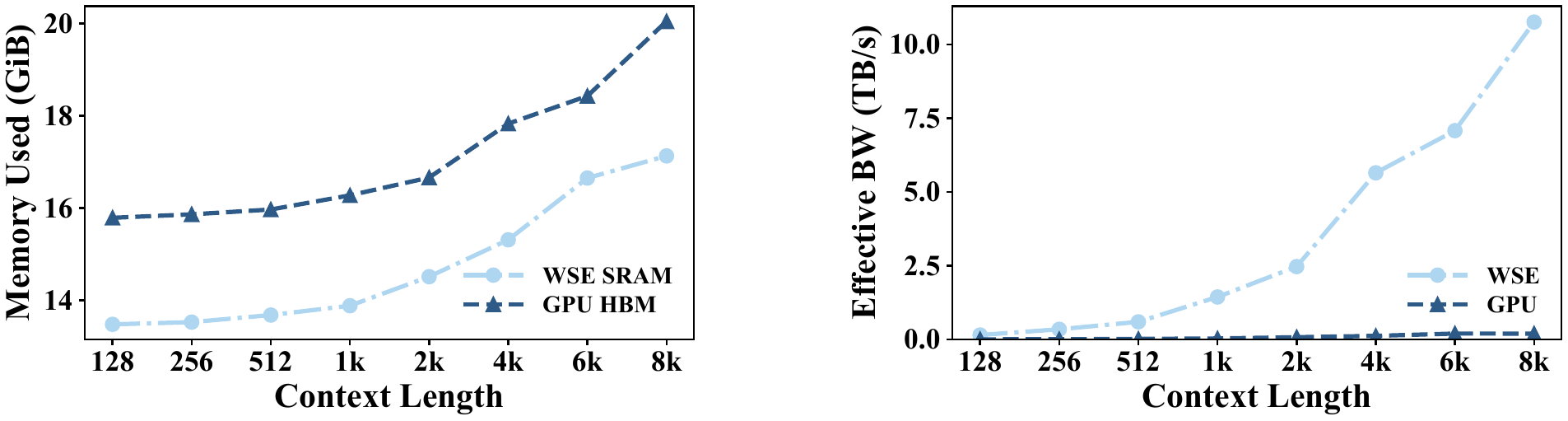}

    \end{minipage}\hfill
           \caption{Memory Footprint and Bandwidth of the WSE and GPU During Decoding.}
\label{fig:decode}
\Description{Memory footprint and effective bandwidth of GPU and WSE execution during decoding.}
\end{figure}
 The WSE can complete a single decode step within tens of microseconds, primarily because of its large-capacity on-chip SRAM and dataflow execution model. As shown in Figure \ref{fig:decode}, GPUs store the KV cache in off-chip HBM, whereas the WSE can place the relevant working set in distributed on-chip SRAM. Because decoding has low arithmetic intensity, its performance is approximately bounded by $AI \times BW$. The high bandwidth of on-chip SRAM therefore raises the WSE's effective performance ceiling while reducing data movement overhead. Meanwhile, the dataflow model can trigger downstream computation as soon as data are available, making it particularly suitable for decoding, which involves limited computation and strong data dependencies. As the context length and KV-cache traffic increase, GPU HBM approaches its bandwidth limit more quickly, causing decode latency to increase significantly. In contrast, the WSE maintains relatively stable latency for each decode step.

\begin{figure}[htbp]

    \centering
    \begin{minipage}[t]{\linewidth}
        \centering
        \includegraphics[width=\linewidth]{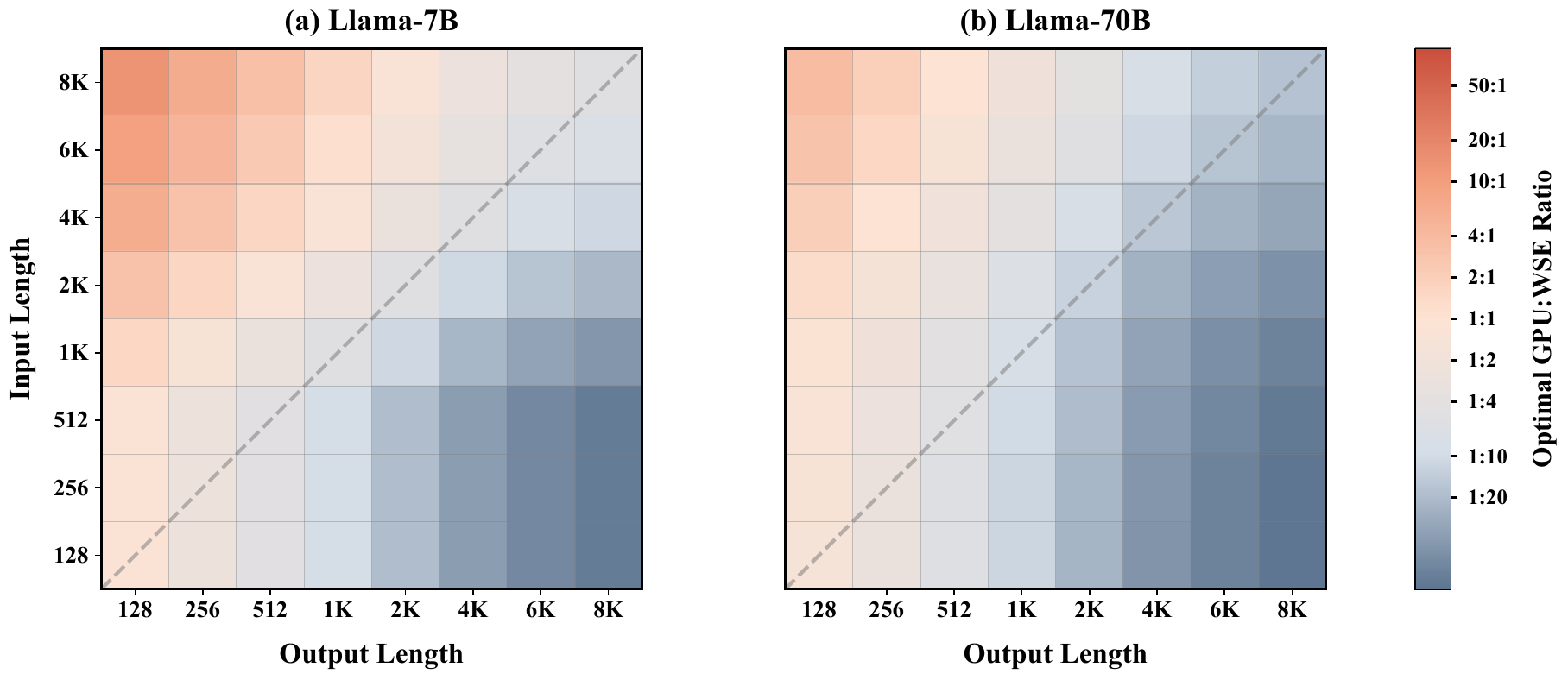}

    \end{minipage}\hfill
           \caption{Optimal GPU--WSE Resource Ratios for Different Request Shapes.}
\label{fig:op}
\Description{Heatmaps show optimal GPU-to-WSE resource ratios for different input and output request lengths.}
\end{figure}
\begin{figure}[htbp]

    \centering
    \begin{minipage}[t]{\linewidth}
        \centering
        \includegraphics[width=\linewidth]{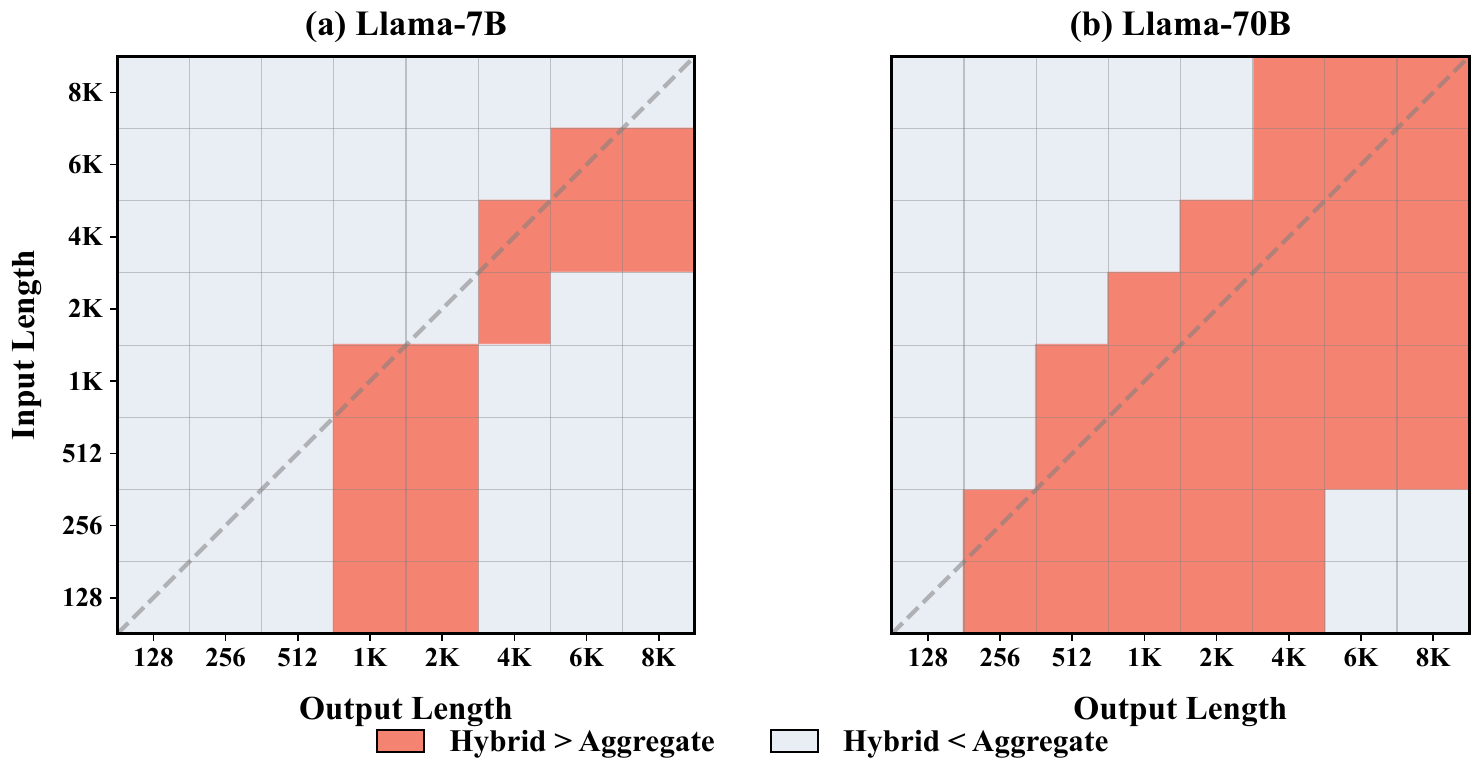}

    \end{minipage}\hfill
           \caption{Hybrid Execution Throughput vs. Aggregate Standalone Throughput.}
\label{fig:hyb}
\Description{Heatmaps show where a fixed heterogeneous configuration improves on aggregate standalone throughput across request shapes.}
\end{figure}

After observing the architectural complementarity between GPUs and WSEs, we further investigate how this complementarity can be translated into actual performance gains. As shown in Figure \ref{fig:op}, the optimal GPU--WSE hardware ratio varies across request shapes as the relative prefill and decode workloads change. We then fix the hardware ratio to the optimal configuration for the $1\mathrm{K}/1\mathrm{K}$ request shape and apply it to other request shapes, as shown in Figure \ref{fig:hyb}. The results show that a single fixed hardware ratio does not provide performance gains across all request shapes. Compared with the 7B model, GPUs and WSEs exhibit more pronounced performance specialization across the prefill and decode stages for the 70B model, resulting in stronger architectural complementarity and a larger opportunity for performance gains. Therefore, for dynamically changing workloads, architectural complementarity alone does not guarantee performance improvement, and dynamic load balancing remains a key issue.
\Needspace{5\baselineskip}
\subsubsection{Layer-Wise Performance Analysis}\leavevmode\par\nopagebreak[4]
\paragraph*{Application Layer Analysis}
Unlike DP and PP, P/D disaggregation decomposes the workload into sub-tasks with substantially different arithmetic intensities. For decoding, the large on-chip SRAM capacity of the wafer-scale architecture allows model weights and the KV cache to remain resident on chip, while its high on-chip memory bandwidth provides a substantially higher memory roof than GPU HBM. As a result, the WSE can outperform the GPU in the low-arithmetic-intensity decode regime. In contrast, the higher-arithmetic-intensity prefill stage is better able to exploit the GPU's compute capability. P/D disaggregation therefore exposes different arithmetic-intensity regimes to the architectures best suited to them, creating a workload--hardware matching space in which architectural differences can be exploited as complementarity.

\paragraph*{System Layer Analysis}
At the system layer, the compiler must first identify and separate workloads that are suitable for execution on GPUs and WSEs, while also considering working-set size to ensure, whenever possible, that workloads mapped to the WSE and their associated data can remain resident in on-chip SRAM and thereby exploit the high bandwidth of the wafer-scale architecture. Second, Figure~\ref{fig:op} shows that different request shapes require different optimal GPU--WSE resource ratios. The system therefore faces a scheduling space jointly determined by hardware quantity and ratio, request characteristics, and cross-hardware task mapping. As the request distribution and system load change dynamically, the same GPU--WSE configuration may correspond to different optimal workload allocations. The runtime scheduler must therefore dynamically select task routing and determine how much workload should be assigned to GPUs and WSEs. Furthermore, when the dynamic workload does not match the available hardware ratio and leaves one hardware domain underutilized, the system should support work stealing to redirect part of the workload to the non-preferred hardware and mitigate load imbalance. Finally, the communication layer must support efficient cross-domain point-to-point data transfer, together with conversion and redistribution of distributed KV-cache representations between GPUs and WSEs, to enable coordinated execution across the two hardware domains.
\begin{table}[h]
\centering
\caption{Lower bound of average KV-cache bandwidth under steady-state prefill throughput.}
\label{tab:kv_bw}
\begin{tabular}{lccc}
\toprule
\textbf{Model Size} & \textbf{Device} & \textbf{Node} & \textbf{Cluster} \\
\midrule
7B  & 6.55 GB/s  & 26.2 GB/s  & 209.6 GB/s \\
70B & 4.6 GB/s  & 18.8 GB/s  & 150.4 GB/s \\
\bottomrule
\end{tabular}

\end{table}
\paragraph*{Hardware Layer Analysis}
Figures~\ref{fig:op} and~\ref{fig:hyb} show that the GPU--WSE hardware ratio is closely related to workload demand. Therefore, when designing such heterogeneous systems, the hardware configuration should be selected to cover as broad a range of workloads as possible, allowing more workloads to exploit the architectural complementarity between the two platforms. In addition, we model the bandwidth demand based on prefill and use the bandwidth required at peak prefill throughput as the minimum communication bandwidth needed to sustain steady-state execution. Table~\ref{tab:kv_bw} reports the bandwidth requirements at the device, node, and cluster levels under our GPU configuration. Overall, compared with DP and PP, P/D disaggregation imposes significantly higher network bandwidth requirements.

\begin{tcolorbox}[colback=gray!20, 
  colframe=black, 
  boxsep=2pt,        %
  left=2pt,          %
  right=2pt,         %
  top=2pt,           %
  bottom=2pt         %
  ]
\textit{\textbf{Takeaway 3}}: Under P/D disaggregation, GPU--WSE architectural differences become exploitable complementarity when heterogeneous tasks exhibit distinct arithmetic intensities that align with architecture-specific strengths. Realizing this complementarity requires architecture-aware computing resource provisioning and runtime scheduling.
\end{tcolorbox}

\section{Discussion}

Using the aggregate throughput of independently running GPUs and WSEs as an additive baseline, we observe that DP can readily approach this baseline, PP struggles to reach it, and PD disaggregation can even exceed it. This demonstrates that workload characteristics are the key factor determining whether architectural differences can be effectively exploited. DP provides sufficient flexibility along the data dimension and can therefore absorb performance losses caused by architectural differences by redistributing the workload quantity between the two hardware domains. PP, however, imposes numerous constraints, making it difficult to partition the workload in a way that accommodates architectural differences. PD disaggregation exposes two heterogeneous tasks with different arithmetic intensities, creating opportunities to match workload characteristics with the strengths of each hardware architecture and thereby exploit architectural complementarity. Therefore, future GPU--WSE heterogeneous systems should be designed with the characteristics of their target applications in mind to avoid resource underutilization and performance loss.

\begin{tcolorbox}[colback=gray!20, 
  colframe=black, 
  boxsep=2pt,        %
  left=2pt,          %
  right=2pt,         %
  top=2pt,           %
  bottom=2pt         %
  ]
\textit{\textbf{Takeaway 4}}: Architectural heterogeneity pays off only when workload structure provides sufficient flexibility to either absorb architectural asymmetry or align heterogeneous tasks with architecture-specific strengths. The system stack determines whether this potential complementarity can be translated into end-to-end performance gains.
\end{tcolorbox}

\section{Performance Modeling}
Building on the three case studies, we develop a lightweight performance model that captures how workload partitioning, architectural differences, and arithmetic intensity jointly shape GPU--WSE heterogeneous execution.
\subsection{Performance Model}

\paragraph*{System abstraction}
Since all experiments are conducted on combinations of two homogeneous systems, we do not extend our model to a more general multi-subsystem setting. Thus, we consider a heterogeneous system composed of two homogeneous subsystems, denoted as

\[
HE = \{HE^1, HE^2\},
\]

where the \(i\)-th subsystem is represented as
\[
HE^i = (P_i, B_i, C_i, M_i).
\]

Here, \(P_i\), \(B_i\), \(C_i\), and \(M_i\) denote the peak compute capability, memory bandwidth, intra-system communication bandwidth, and memory capacity of subsystem \(i\), respectively. These parameters collectively characterize the fundamental hardware constraints governing execution within each subsystem, including computational capacity, data access capability, and resource scale. In addition, we use \(S\) to denote the effective communication bandwidth between the two subsystems, which captures the system-level constraint on cross-subsystem data exchange during coordinated execution.

\paragraph*{Model overview}
Given the system hardware parameters, we model the performance of the heterogeneous system as
\begin{equation}
\label{eq:performance_model}
y = f_{\{HE, S\}}(w).
\end{equation}
Here, \(HE\) and \(S\) denote fixed system parameters. The workload \(w\) is treated as an independent variable and is partitioned into two components, \(w_1\) and \(w_2\), mapped to different subsystems. 

\paragraph*{Model Details}
The model details are given in Eq.~(\ref{eq:perf})--Eq.~(\ref{4}). 
The formulation consists of three components: the end-to-end performance function, the subsystem-level performance model, and the utilization degradation model.

\begin{equation}
\label{eq:perf}
f=\min\left(
\sum_{i=1}^{2} g_i,\;
v\,I(w)\,S
\right),
\end{equation}

\begin{equation}
\label{eq:gi}
g_i=\min\bigl(
u^i_{P}HE^i_P,\;
u^i_{B}w_{AI_i}HE^i_B,\;
u^i_{C}w_{CI_i}HE^i_C
\bigr),
\end{equation}

\begin{equation}
\label{3}
u^i =
\frac{u^{\mathrm{HO},i}(w_i)}
{1+\eta^i(w_1:w_2)},
\qquad
u^{\mathrm{HO},i}(w_i)\propto w_i,
\end{equation}

\begin{equation}
\label{4}
\begin{aligned}
\eta^i(w_1:w_2)=
\sum_{j=1}^{4}
\alpha_j^i(w_1:w_2)\,
\frac{\left|HE_j^1-HE_j^2\right|}
{\max\left(HE_j^1,HE_j^2\right)},\\
\qquad i\in\{1,2\}.
\end{aligned}
\end{equation}
Eq.~(\ref{eq:perf}) models the overall performance of heterogeneous execution as the minimum between aggregated subsystem computation and inter-subsystem communication. 
The term \(g_i\) denotes the effective performance of subsystem \(i\), while \(vI(w)S\) captures the communication limitation.

Eq.~(\ref{eq:gi}) specifies \(g_i\) using a Roofline-style model, where performance is bounded by the dominant bottleneck among compute, memory bandwidth, and communication bandwidth. The workload characteristics are described by the memory access intensity \(w_{AI_i}\) and the communication intensity \(w_{CI_i}\).

Building on this, Eq.~(\ref{3}) introduces the utilization vector \(u=(u_P,u_B,u_C)\), which characterizes the utilization of system resources. 
In heterogeneous settings, resource asymmetry leads to reduced achievable utilization compared to the homogeneous case \(u^{HO}\). 
This degradation is quantified by Eq.~(\ref{4}), where \(\eta(w)\) captures the degree of asymmetry across resource dimensions and its impact on performance, with \(\alpha_j(w)\) representing the penalty coefficient to imbalance in the \(j\)-th resource dimension.

\subsection{Model Validation}

Due to space constraints, we select two representative experiments for model validation. 
\paragraph*{Data Parallel}

As shown in Fig.~\ref{fig:dp_latency}, as the workload increases, all results exhibit a three-stage, two-threshold performance behavior. We now use our model to explain and characterize this behavior. Ignoring interconnect effects, Eq.~(\ref{eq:perf}) shows that the system performance satisfies $f \approx g_1 + g_2$. When the workload partition is optimal, Eq.~(\ref{4}) gives $\alpha_j(w)=0$, which leads to $\eta(w)=0$. Substituting into Eq.~(\ref{3}), the utilization satisfies $u^i = u^{\mathrm{HO^i}}$. 

For small workloads, $w=(0,w)$, resulting in one subsystem being effectively inactive ($g_j \approx 0$). The system is thus dominated by a single subsystem, i.e., $f \approx g_i$. As the workload exceeds the first threshold, the performance can be approximated as $f \approx g_1 + g_2$, where both subsystems become active ($g_1, g_2 > 0$) but remain under-utilized. In this regime, according to Eq.~(\ref{3}), as $w_1$ and $w_2$ increase, $u^{\mathrm{HO}^i}$ increases, which in turn increases $u^i$. This raises $g_i$ in Eq.~(\ref{eq:gi}) and consequently increases $f$ in Eq.~(\ref{eq:perf}). As the workload increases further, the utilization of each subsystem approaches its maximum, i.e., $u^i$ approaches $u^i_{\max}$ in Eq.~(\ref{3}), and the performance becomes bounded by the aggregate subsystem throughput. From Eq.~(\ref{eq:perf}), we have $f \approx g_1^{\max} + g_2^{\max} = f^{\max}$. At this point, the system throughput reaches a constant value.

Beyond validation, we further evaluate the model's predictive capability under unseen settings by changing both the dataset and the numbers of GPUs and WSEs. These configurations are not used during model calibration, and we apply the model directly without re-calibration. Despite these changes, the model consistently captures the performance trends of heterogeneous DP, achieving a median prediction error of 6.8\% and a 90th-percentile error of 13.4\%.

\paragraph*{PD Disaggregation}
As shown in Fig.~\ref{fig:hyb}, assigning prefill and decode to different subsystems can improve the aggregate throughput. We now use our model to explain this behavior and examine whether it is consistent with the model prediction. Prefill and decode have different arithmetic intensities $w_{AI}$: prefill has a higher arithmetic intensity, whereas decode has a lower arithmetic intensity, causing different limiting terms in Eq.~(\ref{eq:gi}) to become active. On the GPU, decode is bandwidth-bound because the bandwidth term $u^i_{B} w_{AI_i} HE^i_B$ in Eq.~(\ref{eq:gi}) is smaller than the compute term. Mapping decode to the WSE increases this bandwidth term, thereby raising $g_i$ toward a higher bound in Eq.~(\ref{eq:gi}). Conversely, prefill is compute-bound on the WSE, and mapping it to the GPU increases the compute term $u^i_{P}HE^i_P$ in Eq.~(\ref{eq:gi}), thereby also increasing $g_i$. Consequently, the performance $g_i$ of each subsystem increases, leading to a larger $\sum_i g_i$ and, in turn, higher overall performance $f$ in Eq.~(\ref{eq:perf}). This model-based analysis explains the observed throughput improvement: better workload--hardware alignment increases the effective performance $g_i$ of each subsystem, thereby improving the aggregate system performance $f$.

Beyond validation, we further evaluate the model's predictive capability using request shapes that are not included in Fig.~\ref{fig:hyb}. We use the calibrated model to predict heterogeneous P/D performance under these unseen request shapes and compare the predictions with the corresponding measurements. The model achieves a median prediction error of 8.2\% and a 90th-percentile error of 16.4\%, indicating that it can reasonably capture performance variation across previously unseen request compositions.

\section{Related Work}
\textit{Heterogeneous GPU Systems.}
Existing studies mainly focus on heterogeneous clusters composed of different generations of GPUs within a single vendor ecosystem. 
In training, systems such as \cite{DBLP:conf/usenix/ParkYYNLCNC20}, \cite{DBLP:conf/ics/GuoACD25}, and \cite{DBLP:journals/corr/abs-2504-03871} incorporate GPUs with diverse capabilities into a unified training pipeline through non-uniform partitioning of computation and model structure, to mitigate imbalances in compute and memory resources. 
In inference, systems such as \cite{DBLP:conf/asplos/Mei0MYJV25} model GPU and network resources, and optimize model placement and request routing to improve serving performance on heterogeneous clusters. 
In addition, a small number of recent efforts begin to explore multi-vendor GPU heterogeneity. 
For example, \cite{DBLP:journals/corr/abs-2601-22585} enables efficient communication across GPUs from different vendors, while \cite{DBLP:journals/corr/abs-2509-17542} explores LLM serving across multi-vendor GPU environments.

\textit{Roofline-based Performance Models for Distributed Systems.} Many works extend the Roofline model to modern multi-device settings. \cite{DBLP:conf/sc/0006ALMFBONW024} generalizes Roofline from single-program analysis to end-to-end workflow analysis, identifying node-level, system-level, preprocessing-level, and control-flow bottlenecks in scientific workflows on HPC systems. \cite{DBLP:conf/hpcasia/CardwellS19}, \cite{DBLP:journals/corr/abs-2209-01368} further extend Roofline to distributed-memory HPC and distributed systems by incorporating communication into the model, thereby jointly characterizing performance bounds imposed by multi-resources. \cite{DBLP:conf/springsim/TangHB15} extends roofline-style modeling to CPU+GPU heterogeneous systems, focusing on the performance upper bound under a given workload partition.

\section{Conclusion}
This work studies heterogeneous GPU--WSE execution across the application, system, and architecture layers. Through DP, PP, and P/D disaggregation, we show when architectural differences become performance obstacles or exploitable complementarity, governed primarily by workload partitionability and task--hardware affinity. We further develop a unified performance model to capture these effects and guide heterogeneous system design and scheduling.

\bibliographystyle{ACM-Reference-Format}
\bibliography{references}

\end{document}